\documentstyle[aps,prl,preprint,epsf]{revtex}
\input psfig

\newcounter{saveeqn}

\newcounter{savefig}

\begin{document}

\title{Structural evidence for the \\ continuity of liquid and glassy
water}

\author{Francis~W. Starr$^\ast$, Marie-Claire
Bellissent-Funel$^\dagger$, and H.~Eugene Stanley$^\ast$}

\address{$^\ast$Center for Polymer Studies, Center for Computational
Science, and Department of Physics, Boston University, Boston, MA
02215 USA}

\address{$^\dagger$Laboratoire L\'eon Brillouin (CEA-CNRS), CEA/Saclay, 91191
Gif-sur-Yvette, Cedex, France}

\bigskip

\maketitle

{\bf An open question is whether the liquid and glassy phases of water
are thermodynamically distinct or continuous~\cite{debenedetti}.  Here
we address this question using molecular dynamics simulations in
comparison with neutron scattering experiments to study the effect of
temperature and pressure on the local structure of liquid water.  From
both simulations and experiments, we find that the liquid structure at
high pressure is nearly independent of temperature, and remarkably
similar to the known structure of the high-density amorphous ice (HDA).
Further at low pressure, the liquid structure appears to approach the
experimentally-measured structure of low-density amorphous ice (LDA)as
temperature decreases.  These results are consistent with continuity
between the liquid and glassy phases of H$_2$O.}

The structure of liquid water has been well-studied at ambient pressure
by a variety of experimental and simulation techniques.  It has been
recognized that each water molecule is typically hydrogen bonded to four
neighboring molecules in a tetrahedral arrangement, leading to an open
bond network that can account for a variety of the known anomalies of
water \cite{debenedetti}.  More recently, the effect of pressure on both
the structure and the hydrogen bond network of liquid water has been
studied experimentally \cite{wwd82,odg94,mcbf95} and by simulations
using a variety of potentials, including the ST2 potential
\cite{sr74-2,pses1,pses3,spes97}, the MCY potential \cite{ikm81}, the
TIP4P potential \cite{pses3,spes97,tk88,mpc88,tan96}, and the SPC/E
potential \cite{hpss97,bbk97}.  Furthermore, understanding the effects
of pressure may be useful in elucidating the puzzling behavior of liquid
water.

In particular, three competing ``scenarios'' have been hypothesized to
explain the anomalous properties of water: (i) the existence of a
spinodal bounding the stability of the liquid in the superheated,
stretched, and supercooled states \cite{sa76,speedy82}; (ii) the
existence of a liquid-liquid transition line between two liquid phases
differing in density \cite{pses1,pses3,spes97,tan96,rpd96,mc98};
(iii) a singularity-free scenario in which the anomalies are related to
the presence of low-density and low-entropy structural heterogeneities
\cite{sastry96}.

Here, we present molecular dynamics (MD) simulations (Table
\ref{state-points}) of a comparatively large system of 8000 molecules
interacting via the extended simple point charge (SPC/E) pair potential
\cite{spce}.  We find remarkable agreement with neutron scattering
studies of the effect of pressure on the structure of liquid D$_2$O
\cite{mcbf95}, indicating that the SPC/E potential reproduces many
structural changes in the liquid over a wide range of temperature and
pressure.  By comparing the simulated pair correlation functions and
structure factor with our experimental data, we find that the structure
of the supercooled liquid at low pressure resembles the structure of
low-density amorphous (LDA) ice.  At high pressure, we find that the
structure of the liquid appears independent of temperature and is nearly
indistinguishable from that of high-density amorphous (HDA) ice.  The
combined results at high and low pressure appear consistent with
continuity between the liquid and glassy states of water and also with a
possible liquid-liquid transition in the supercooled region of the
liquid terminated by a second critical point.

We analyze the structures found in our simulations by calculating the
atomic radial distribution functions (RDF) $g_{OO}(r)$, $g_{OH}(r)$, and
$g_{HH}(r)$.  To compare the RDFs with neutron scattering measurements,
we form the weighted sum

\begin{equation}
h(r) \equiv 4 \pi \rho r \left[ w_1 g_{OO}(r) + w_2 g_{OH}(r) + w_3
g_{HH}(r) - 1 \right],
\label{h(r)-eq}
\end{equation}

\noindent where the weighting factors $w_i$ are selected to coincide
with experimental measurements of D$_2$O.  Experimentally, $h(r)$ is
obtained by Fourier transformation of the molecular structure factor.
The dominant contributions to $h(r)$ are the H-H and O-H (or D-D and O-D
for D$_2$O) spatial correlations, so $h(r)$ includes relatively little
contribution from oxygen-oxygen correlations.  Fig.~\ref{fig1}a shows
$h(r)$ at two of the five temperatures simulated and also compares with
experimental data.  The peaks centered at 1.8~\AA, 2.3~\AA, and 2.8~\AA\
correspond to the O-H, H-H, and O-O intermolecular distances in the
hydrogen-bonded configuration, respectively.  While the magnitudes of
these peaks change slightly, their ubiquity demonstrates the stability
of the first neighbor ordering -- namely that each molecule is typically
surrounded by four molecules in a tetrahedral configuration.

We find that the peak at 3.3~\AA\ of $h(r)$ becomes more pronounced as
$P$ increases.  Examination of the individual RDFs shows that the
increase at 3.3~\AA\ can be attributed to changes in $g_{OH}(r)$.  In
addition, $g_{OO}(r)$ shows a pronounced increase at 3.3~\AA\ under
pressure (Fig.~\ref{fig1}b), but cannot account for the changes in
$h(r)$, as the $g_{OO}(r)$ weighting factor in Eq.~(\ref{h(r)-eq}) is
small~\cite{wwd82,odg94,mcbf95}.  The growth at 3.3~\AA\ in $g_{OO}(r)$
indicates that the liquid locally has the structure of an
interpenetrating tetrahedral network similar to ice VI and VII, the ice
polymorphs close to the high pressure liquid, and can also be associated
with the formation of clusters with structure similar to HDA
\cite{sasai}.

To directly compare with experimental measurements, we calculate the
molecular structure factor $S_M(q) = S_M^{intra}(q) + S_M^{inter}(q)$,
where $S_M^{intra}(q)$ and $S_M^{inter}(q)$ are the intramolecular and
intermolecular contributions to $S_M(q)$, respectively.  We calculate
$S_M^{intra}(q)$ explicitly, as described in ref.~\cite{mcbf91}, and
Fourier transform $h(r)$ to obtain $S_M^{inter}(q)$.  We find striking
agreement between simulated and the experimental values of $S_M(q)$
(Fig.~\ref{fig1}c).  In particular, the value $q_0 = q_0(P,T)$ of the
first peak of $S_M(q)$ shifts to larger $q$ values as $P$ increases
(Fig.~\ref{q0-fig}).  We note the feature that as temperature is
lowered: (i) for $P=600$ MPa, $q_0$ approaches the experimental $q_0$
value of HDA ice (2.20~\AA$^{-1}$), (ii) For $P=0.1$ MPa and $-200$ MPa,
$q_0$ approaches the experimental $q_0$ value of LDA ice
(1.69~\AA$^{-1}$).  Indeed we find the high-pressure liquid structure
resembles that of HDA ice (Fig.~\ref{HDA-LDA}a), and the low-pressure
liquid structure that of LDA ice (Fig.~\ref{HDA-LDA}b)
\cite{pses1,spes97}.

We also study the structure of the glassy phase and find that the
structure of the simulated glasses strongly resembles the
experimentally-measured structure of HDA and LDA solid water
(Fig.~\ref{HDA-LDA}, c and d), suggesting that the simulated glassy
state points are the analogues of HDA and LDA solid water for SPC/E.
Furthermore, by compression of the simulated LDA-like system, we are
able to reversibly transform the structure to the HDA-like system,
reminiscent of the experimentally observed reversible first-order
transition of LDA to HDA under pressure \cite{mishima94}.

These results support the postulated continuity between atmospheric
pressure water and LDA ice, based on measurements of the free energy
\cite{speedy96} and dielectric relaxation time \cite{johari96} at 1
atm.  Our results are also consistent with continuity between high
pressure liquid water and HDA ice \cite{mcbf95,pses1,pses3,spes97}.
Combined with the experimentally-detected first-order transition between
HDA and LDA, our results are consistent with the presence the postulated
second critical point in the supercooled region of the phase diagram.

\bigskip\bigskip\bigskip
\noindent{\bf\sf Methods}\\ {\small We equilibrate systems of 8000
molecules interacting via the SPC/E potential to a constant temperature
and pressure by monitoring the evolution of the density and internal
energy.  The SPC/E model treats water as a rigid molecule consisting of
three point charges located at the atomic centers of the oxygen and
hydrogen which have an OH distance of 1.0 \AA\ and HOH angle of
109.47$^\circ$, the tetrahedral angle.  Each hydrogen carries a charge
$q_H = 0.4238e$ and the oxygen carries a charge $q_O = -2q_H$, where $e$
is the magnitude of the electron charge.  In addition, the oxygen atoms
of separate molecules interact via a Lennard-Jones potential with
parameters $\sigma = 3.166$ \AA\ and $\epsilon = 0.6502$ kJ/mol.  We
adjust the temperature and pressure via the methods of Berendsen
\protect\cite{ber-84} and we account for the long-range Coulomb
interactions using the reaction field technique with a cutoff of 0.79 nm
\protect\cite{steinhauser}. The equations of motion evolve by the SHAKE
algorithm \protect\cite{SHAKE} with a time step of 1 fs.  We typically
simulate each state point using eight processors in parallel.  We obtain
simulation speeds of approximately $15 \mu s$ per particle per update.
The total simulation time is about 1000 CPU-days.}

\noindent {\bf Acknowledgments:} We thank C.A. Angell, J.~Nielsen,
C.~Roberts, S.~Sastry, F. Sciortino, and J.~Teixeira for helpful
discussions, and especially S.~Harrington for his significant
contributions to the early stages of this work.  We are grateful to the
Center for Computational Science at Boston University for extensive use
of the 192-processor SGI/Cray Origin 2000 supercomputer.  FWS is
supported by a NSF graduate fellowship.  The Center for Polymer Studies
is supported by NSF Grant CH9728854.

\begin{table}
\caption{Summary of results from the 24 liquid and 2 glassy state points
simulated.  The liquid state points simulated at positive pressure
correspond roughly to temperatures and pressures studied in the
experiments of ref.~\protect\cite{mcbf95}.  To facilitate comparison
between simulations and experiments, we define $\Delta T \equiv T -
T_{MD}$, the temperature relative to that of the 1 atm temperature of
maximum density $T_{MD}$.  At atmospheric pressure, the SPC/E potential
displays a density maximum at $T_{MD}^{SPC/E} \approx
245$~K~\protect\cite{hpss97,bbk97,bc94}.  Similarly, experimental
temperatures reported relative to the 1 atm $T_{MD}$ of D$_2$O, 284~K.
All state points are liquid, except $\Delta T = -145~^\circ$C, which is
glassy.  We obtain state points at $\Delta T = -145~^\circ$C and $P =
0.1$ and $600$~MPa by quenching a configuration of the supercooled
liquid state points at $\Delta T = -30~^\circ$C.  The configurations
analyzed at $\Delta T = -145~^\circ$C are not ``equilibrated'', as the
required simulation time far exceeds the computational resources
available; rather, they correspond to a glassy state quenched from the
supercooled liquid.  We consider these state points to be glassy since
previous simulations indicate that SPC/E approaches a glass for $\Delta
T < -50~^\circ$C \protect\cite{bc94,sciortino-and-me}.  At negative
pressures, smaller systems may not reproduce cavitation events that we
observe in larger systems.  Thus it is important that we consider the
large 8000 molecule systems for $P = -200$ MPa.  We also simulated one
system of 64,000 molecules, but did not observe any significant
differences in structure or cavitation events.}
\medskip
\begin{tabular}{cccccc}
$P$ (MPa) & $\Delta T$ ($^\circ$C) & $\rho$ (g/cm$^3$) & $U$ (kJ/mol) && 
Equilibration Time (ps) \\
\tableline
600 & $ 35  $  & $1.1930 \pm 0.0003$ & -47.65 && 1100 \\
    & $ 10  $  & $1.2041 \pm 0.0004$ & -49.02 && 1300 \\
    & $ -10 $  & $1.2139 \pm 0.0005$ & -50.22 && 2000 \\
    & $ -30 $  & $1.2236 \pm 0.0003$ & -51.43 && 3000 \\
    & $ -145 $ & $1.250  \pm 0.008$  & -55.3  && 4000 \\
465 & $ 35  $  & $1.1616 \pm 0.0004$ & -47.61 && 1100 \\
    & $ 10  $  & $1.1723 \pm 0.0005$ & -49.05 && 1300 \\
    & $ -10 $  & $1.1803 \pm 0.0005$ & -50.24 && 1800 \\
    & $ -30 $  & $1.1899 \pm 0.0004$ & -51.49 && 2500 \\
260 & $ 35  $  & $1.1060 \pm 0.0004$ & -47.59 && 550 \\
    & $ 10  $  & $1.1145 \pm 0.0006$ & -49.09 && 750 \\
    & $ -10 $  & $1.1201 \pm 0.0005$ & -50.33 && 1500 \\
    & $ -30 $  & $1.1224 \pm 0.0005$ & -51.62 && 2100 \\
100 & $ 35  $  & $1.0514 \pm 0.0004$ & -47.51 && 600 \\
    & $ 10  $  & $1.0570 \pm 0.0005$ & -49.11 && 800 \\
    & $ -10 $  & $1.0555 \pm 0.0003$ & -50.48 && 1500 \\
    & $ -30 $  & $1.0513 \pm 0.0006$ & -51.85 && 2100 \\
0.1 & $ 35  $  & $1.0132 \pm 0.0003$ & -47.42 && 600 \\
    & $ 10  $  & $1.0158 \pm 0.0004$ & -49.12 && 800 \\
    & $ -10 $  & $1.0132 \pm 0.0003$ & -50.56 && 1500 \\
    & $ -30 $  & $1.0046 \pm 0.0006$ & -52.15 && 3000 \\
    & $ -145 $ & $1.022  \pm 0.009$  & -56.1  && 4000 \\
-200 & 35      & $0.9064 \pm 0.0003$ & -46.73 && 600 \\
    & $ 10  $  & $0.9212 \pm 0.0005$ & -48.83 && 800 \\
    & $ -10 $  & $0.9245 \pm 0.0003$ & -50.59 && 1500 \\
    & $ -30 $  & $0.9283 \pm 0.0004$ & -52.12 && 3000 \\
\end{tabular}
\label{state-points}
\end{table}

\newbox\figa
\newbox\figb
\newbox\figc
\setbox\figa=\psfig{figure=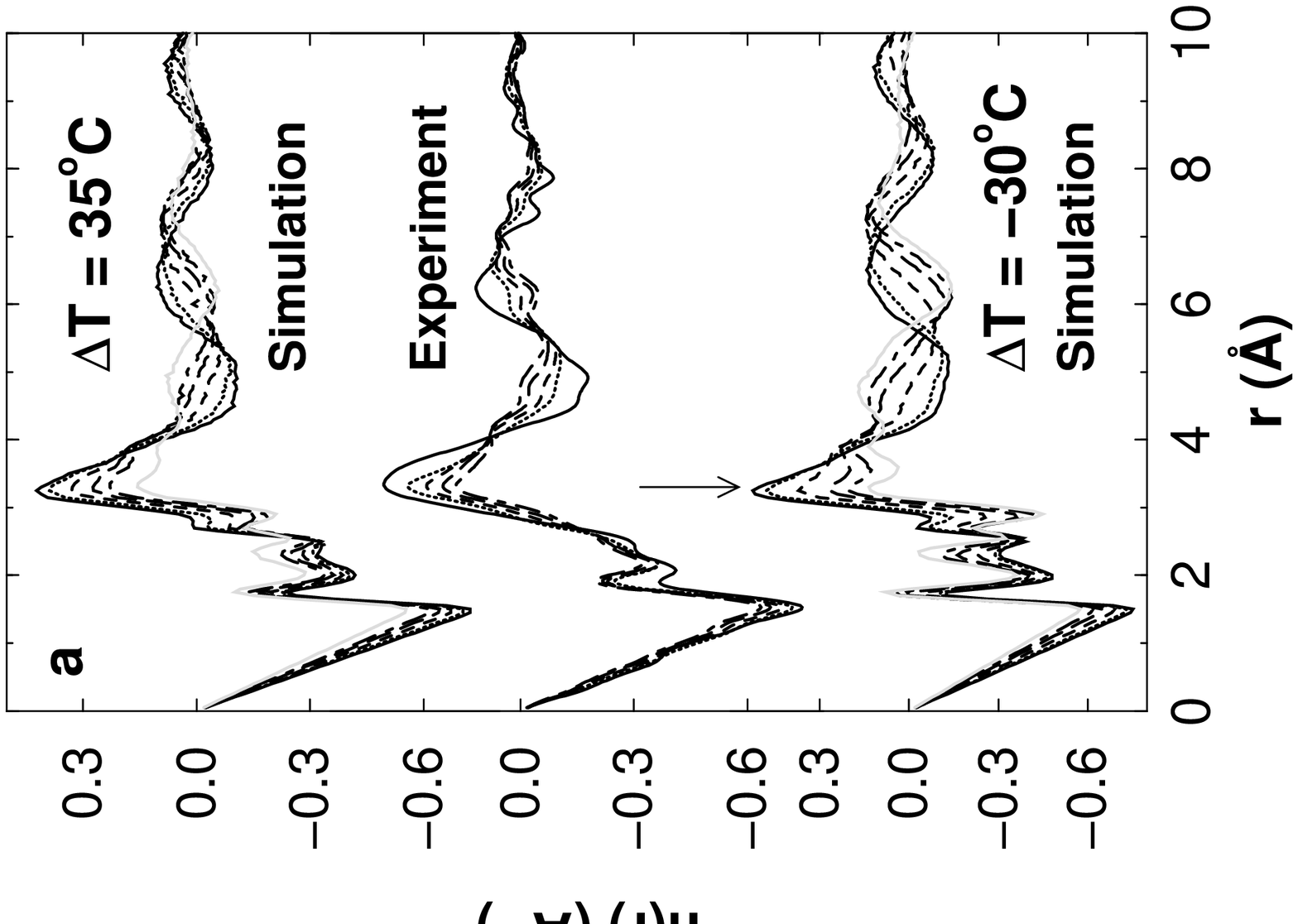,width=10cm,angle=-90}
\setbox\figb=\psfig{figure=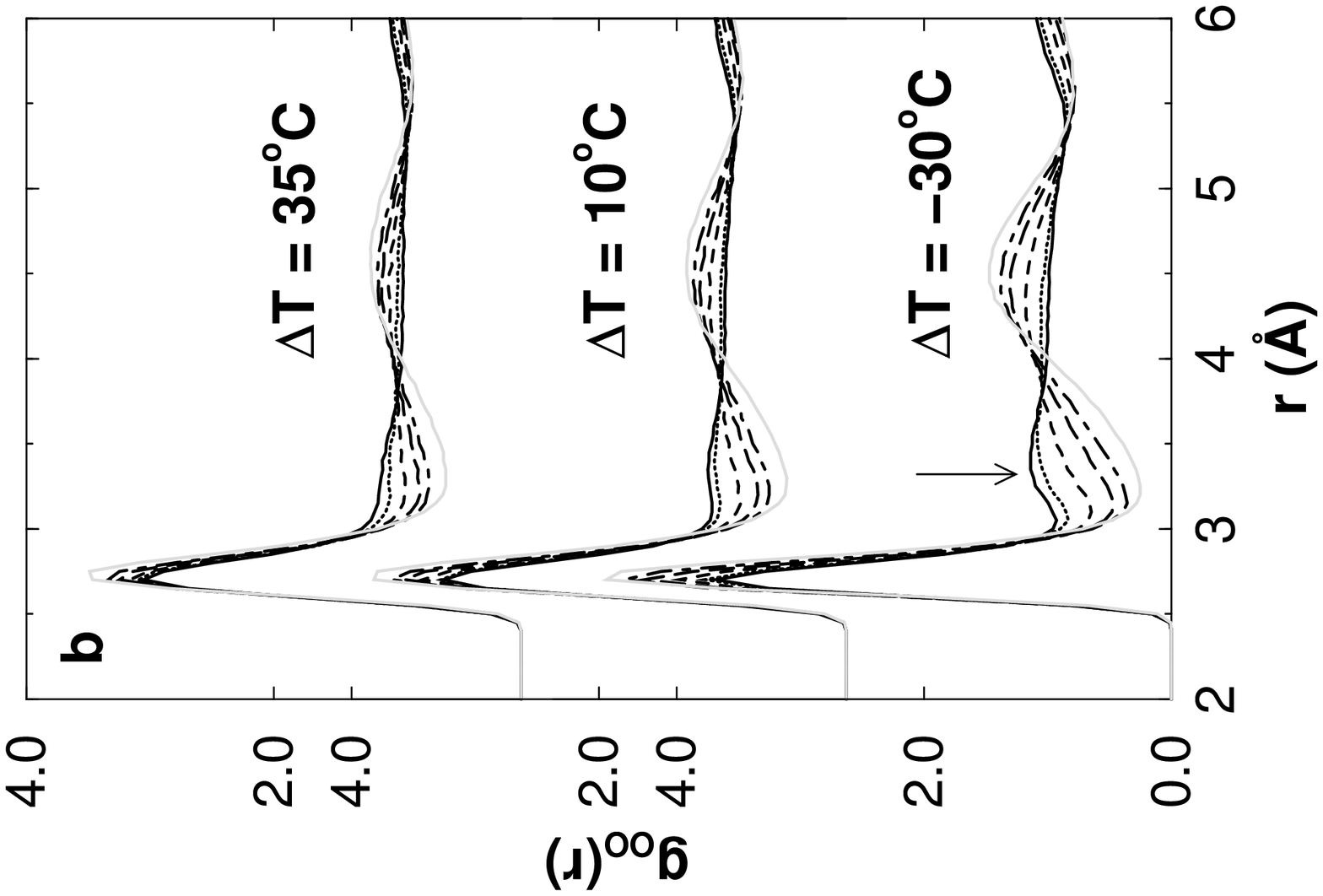,width=10cm,angle=-90}
\setbox\figc=\psfig{figure=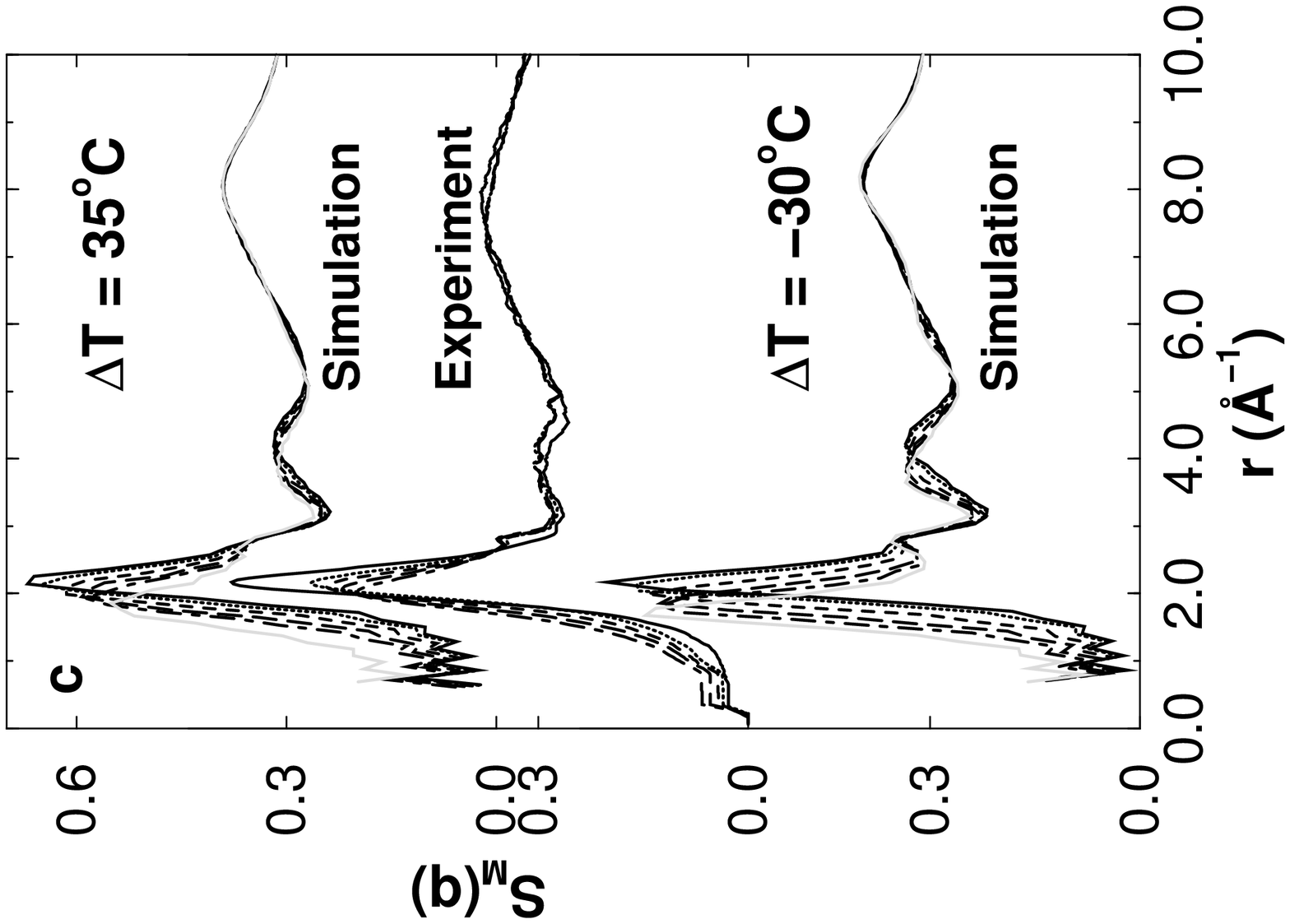,width=10cm,angle=-90}
\begin{figure*}[htbp]
\begin{center}
\leavevmode
\centerline{\box\figa}
\newpage\centerline{\box\figb}
\newpage\centerline{\box\figc}
\narrowtext
\caption{Structure of liquid water comparing results from
experiments~\protect\cite{mcbf95} and the present simulations.  Each set
of curves can be identified as follows (reading from top to bottom at
the location of the arrow): $P = 600$~MPa, 465~MPa, 260~MPa, 100~MPa,
0.1~MPa, and -200~MPa.  {\bf a,} The pair correlation function $h(r)$
for two of five temperatures studied.  Pressures are the same for the
experiments and simulations, with the exception that no experiment was
possible at $P = -200$~MPa.  Note the pronounced increase in the
3.3~\AA\ peak (arrow) when pressure is increased.  To facilitate
comparison with experiments, the simulation temperature is reported
relative to the $T_{MD}$ of the SPC/E potential at atmospheric pressure.
Similarly, the experimental data are reported relative to the $T_{MD}$
of D$_2$O at atmospheric pressure.  Isotopic substitution is reflected
by the weighting factors used to calculate $h(r)$, given by
equation~(\ref{h(r)-eq}).  Substitution of deuterium for hydrogen is
expected to have little effect on the individual RDFs.  To compare with
experimental measurements, we use weighting factors for D$_2$O, given by
$w_1 = 0.092$, $w_2 = 0.422$, and $w_3 = 0.486$.  {\bf b,} The pair
correlation function $g_{OO}(r)$ for three of five temperatures studied.
Note the pronounced increase in the 3.3 \AA\ peak (arrow) when pressure
is increased.  Simulations of the ST2 and MCY potential do not display a
peak at 3.3~\AA\ in $g_{OO}(r)$, characteristic of the interpenetrating
tetrahedral structure expected under high pressure
\protect\cite{sr74-2,ikm81}. {\bf c,} The molecular structure factor
$S_M(q)$, calculated from the Fourier transform of $h(r)$
(Fig.~\ref{fig1}a).  Looking at the first peak in $S_M(q)$, the curves
are identified as described above.  Note the shift in the first peak
when pressure is increased.}
\label{fig1}
\end{center}
\end{figure*}

\newbox\figa
\setbox\figa=\psfig{figure=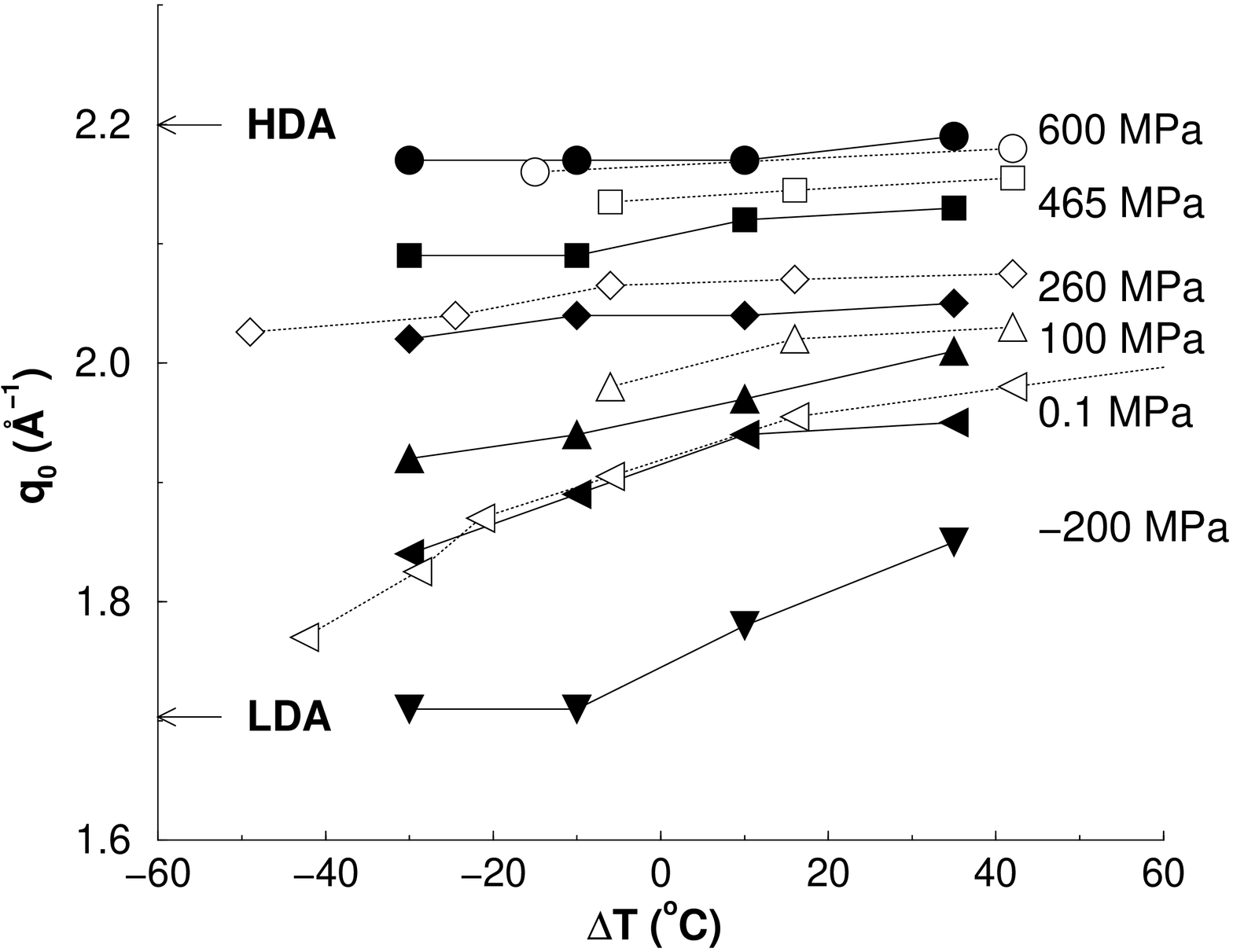,width=10cm,angle=-90}
\begin{figure*}[htbp]
\begin{center}
\leavevmode
\centerline{\box\figa}
\narrowtext
\caption{The value $q_0$ of the first peak of the structure factor from
Fig.~\ref{fig1}c.  At low T, both the simulated (filled symbols) and
experimental (open symbols)~\protect\cite{mcbf95} value of $q_0$ tend
toward the values for the two amorphous forms of water, HDA and LDA.}
\label{q0-fig}
\end{center}
\end{figure*}

\newbox\figa
\setbox\figa=\psfig{figure=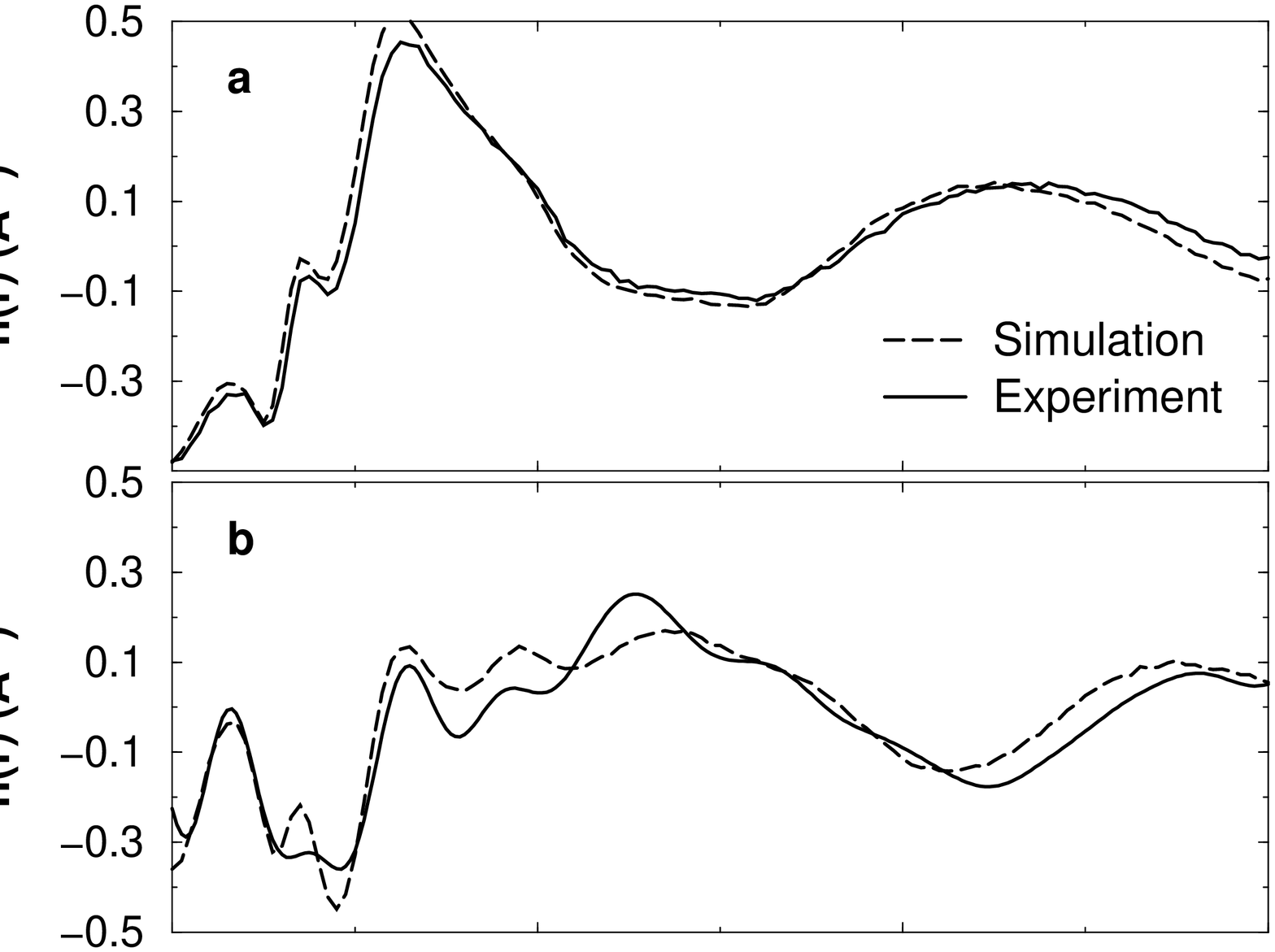,width=10cm,angle=-90}
\newbox\figb
\setbox\figb=\psfig{figure=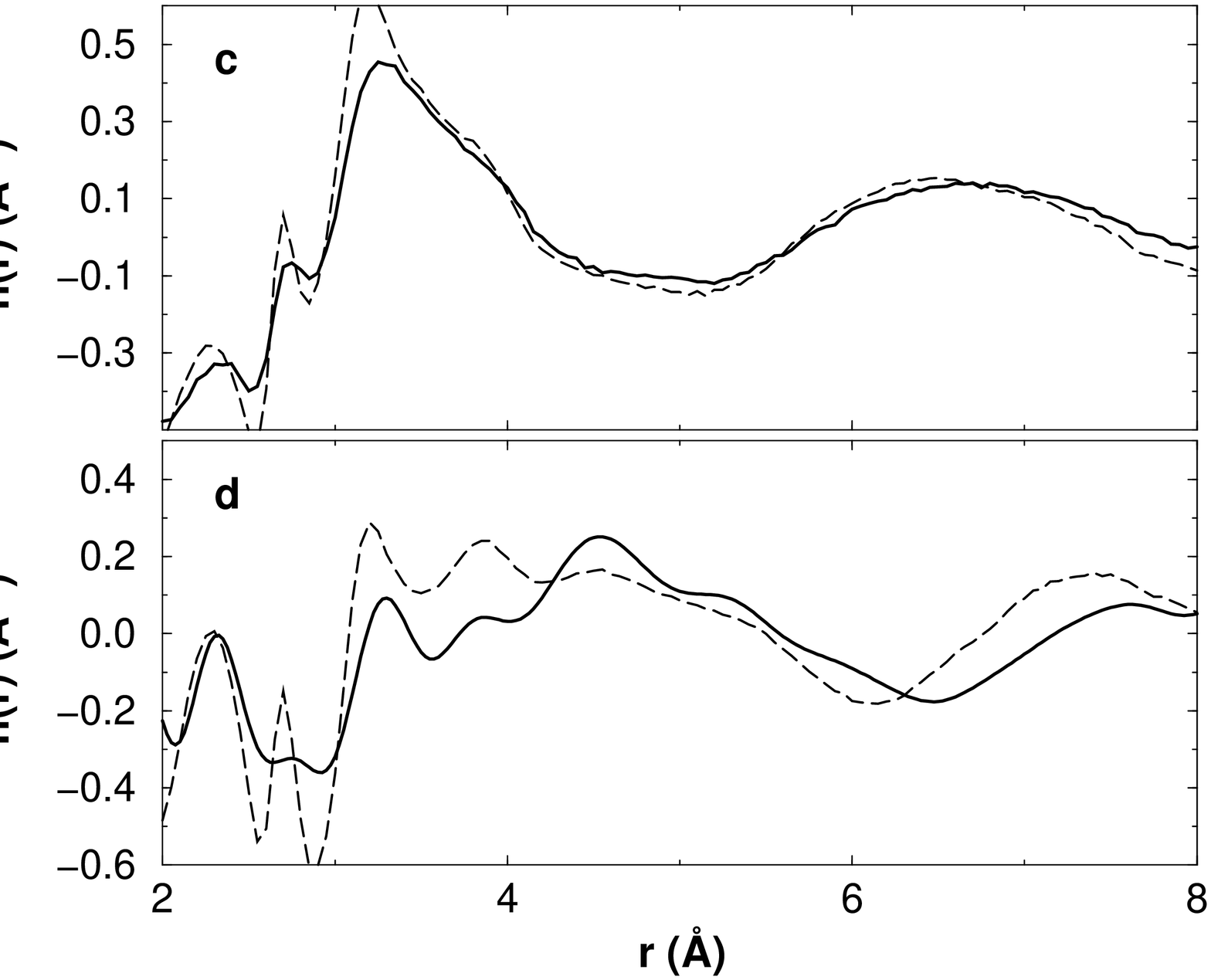,width=10cm,angle=-90}
\begin{figure*}[htbp]
\begin{center}
\leavevmode
\centerline{\box\figa}
\centerline{\box\figb}
\caption{{\bf a,} Comparison of the structure of the supercooled liquid
at $\Delta T = -30~^\circ$C and $P=600$~MPa with the experimentally
measured structure of HDA solid water from ref.~\protect\cite{bftb87}.
{\bf b,} Comparison of the supercooled liquid structure at $\Delta T =
-30~^\circ$C and $P=-200$~MPa with LDA solid water
\protect\cite{bftb87}.  {\bf c,} Comparison of the structure of the
glassy simulation at $\Delta T = -145~^\circ$C and $P=600$~MPa with the
experimentally measured structure of HDA solid water from
ref.~\protect\cite{bftb87}.  {\bf d,} Comparison of the glassy
simulation at $\Delta T = -145~^\circ$C and atmospheric pressure with
LDA solid water \protect\cite{bftb87}.  While negative pressure is
necessary to observe LDA-like structure in the simulations for $\Delta T
= -30~^\circ$C, atmospheric pressure is sufficient to observe LDA-like
structure at $\Delta T = -145~^\circ$C.}
\label{HDA-LDA}
\end{center}
\end{figure*}

\end{document}